# Theoretical evaluation of $[V^{IV}(\alpha\text{-}C_3S_5)_3]^{2-}$ as nuclear-spin sensitive single-molecule spin transistor.

*S. Cardona-Serra*[1,2], *A. Gaita-Ariño*[1], *M. Stamenova*[2] *and S. Sanvito*[2].

[1] Instituto de Ciencia Molecular, Universitat de València, España.

[2] School of Physics, AMBER and CRANN Institute, Trinity College, Dublin 2, Ireland.

**Corresponding Author**

E-mail: salvador.cardona@uv.es

In a straightforward application of molecular nanospintronics to quantum computing, single-molecule spin transistors can be used to measure and control nuclear spin qubits. A jump in the conductance occurs when the electronic spin inverts its polarization, and this happens at a so-called anticrossing between energy levels, which in turn only takes place at a specific magnetic field determined by the nuclear spin state. So far, this procedure has only been implemented for the terbium(III) bis(phthalocyaninato) complex. Here we explore theoretically whether a similar behavior is expected for a highly stable molecular spin qubit, the vanadium tris-dithiolate complex $[V^{IV}(\alpha\text{-}C_3S_5)_3]^{2-}$. We consider such molecule sandwiched into a two-terminal device and determine the spin-dependent conductance. We verify that the transport channel at minimal bias voltage does not overlap with the occupied spin orbitals, indicating that the spin states may survive in the conduction regime. We estimate some physical parameters to guide the experiments, and verify the robustness of the theoretical methodology by applying it to two chemically related vanadium complexes.

**TOC GRAPHICS**

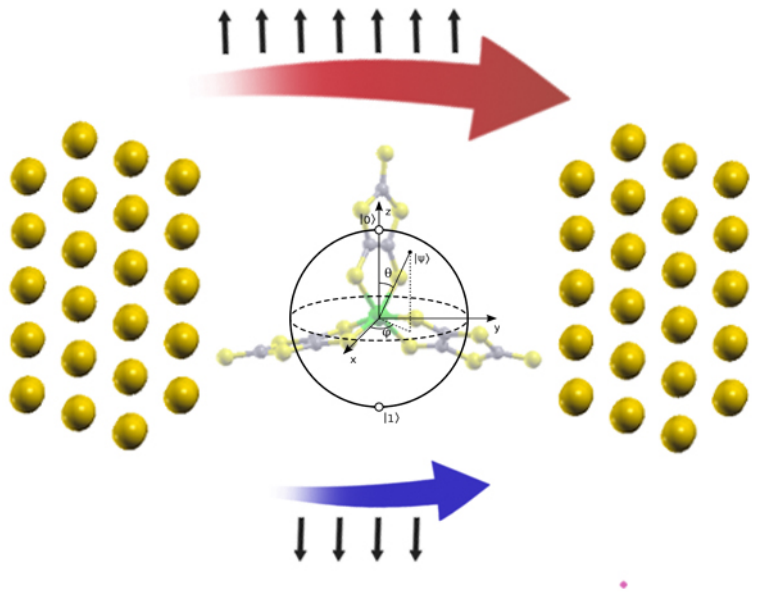

Molecular spintronics is an emerging branch of nanotechnology that builds on previous knowledge acquired in its inorganic counterpart, inorganic spintronics, and adds the infinite possibilities of tailoring and tuning the electronic properties of molecules.[1,2,3] The field proposes new materials to the data storage and data processing industries but also aims at offering novel possibilities to quantum technologies. In a seminal experiment carried out in Wernsdorfer's group,[4,5] the read-out of a two-qubit quantum state was performed by the measurement of the conductance of a single terbium bisphthalocyaninato ($[TbPc_2]^-$) molecule, which acted as a hyperfine-controlled single-molecule spin transistor. In this experiment, the two qubits reside in the $I$ = 3/2 nuclear spin quadruplet of a single $^{159}Tb^{3+}$ ion, $J$ = 6 with a ground doublet $M_J$ = ±6. The nuclear spin states are read out by a spin-dependent current: the conductance exhibits a jump at an anticrossing between electronuclear spin levels, at a magnetic field determined by the nuclear spin projection $M_I$ (see Figure 1, bottom panel). This effect is only measurable by working at low temperature, where the molecule is in a fully polarized electronic spin state. Recently, coherent quantum manipulation has been performed by the hyperfine Stark effect to implement a minimal four-level Grover algorithm in a single molecule using this approach.[6] While these experiments so far have been implemented only with $[TbPc_2]^-$, chemistry could offer a palette of molecular structures matching or improving the possibilities of manipulation. There are two necessary requirements: firstly, anticrossings between electronuclear spin states should exist and be accessible, and secondly, the transport channel needs to be orthogonal to spin orbitals to allow the survival of the electronuclear spin states in the transport regime. In addition, further characteristics are desirable, such as a slow spontaneous relaxation of the electronic spin to allow longer operation times, a higher nuclear spin to increase the number of qubits ($n$ qubits require $2^n$ states), and a large quantum tunneling to improve the Landau-Zener transition probability at the anticrossings and thus the operational speed.

A theoretical study is useful to explore the wide variety of chemical systems that can in principle fulfill these requisites and thus be candidates for such kind of experiments. Among lanthanide-based single ion magnets (SIMs) but beyond the phthalocyaninato sandwiches used in the original experiments, other obvious candidates are polyoxometalate complexes[7,8,9] because of their electronic versatility,[10,11] and in particular $DyW_{30}$, which has been experimentally proven to function as a single-molecule rectifier.[12] We did not choose this kind of systems for the present study because our electronic structure method of choice, local and semi-local density functional theory, does not provide a good description of ions containing 4*f* electrons (see SI for a broader discussion on the process of selection of the molecule to study). A second chemical family of potential interest is the recently investigated and smaller family of transition-metal based molecular spin qubits.[13,14,15,16] Among them, we decided to start our exploratory search with the vanadium complex $[V(\alpha\text{-}C_3S_5)_3]^{2-}$ (figure 1): [13] This presents a high nuclear spin octuplet $I$=7/2, meaning its 2x2x2 states give rise to 3 qubits for each single molecule.[9] This complex presents a tunneling splitting that is at least three orders of magnitude above the 1 $\mu$K found in $[TbPc_2]^-$, improving the operational speed (see Supplementary Information). Furthermore, $[V(\alpha\text{-}C_3S_5)_3]^{2-}$ has the potential to be attached to gold through its sulfur terminal atoms, and it presents a long relaxation time that increases by cooling (the rapidly increasing times $T_1$ = 0.05, 1, 20 ms are achieved just by lowering the temperature from 40 K to 20 K and 10 K).[13] Moreover, because of its small size it is computationally inexpensive to deal with. Although its vanadyl analogue, $[VO(\alpha\text{-}C_3S_5)_2]^{2-}$, is expected to present a much more convenient magnetic field separation between the anticrossings, we chose $[V(\alpha\text{-}C_3S_5)_3]^{2-}$ since preliminary calculations indicated that it is more favorable from the point of view of spin-dependent conductance (see Supplementary Information).

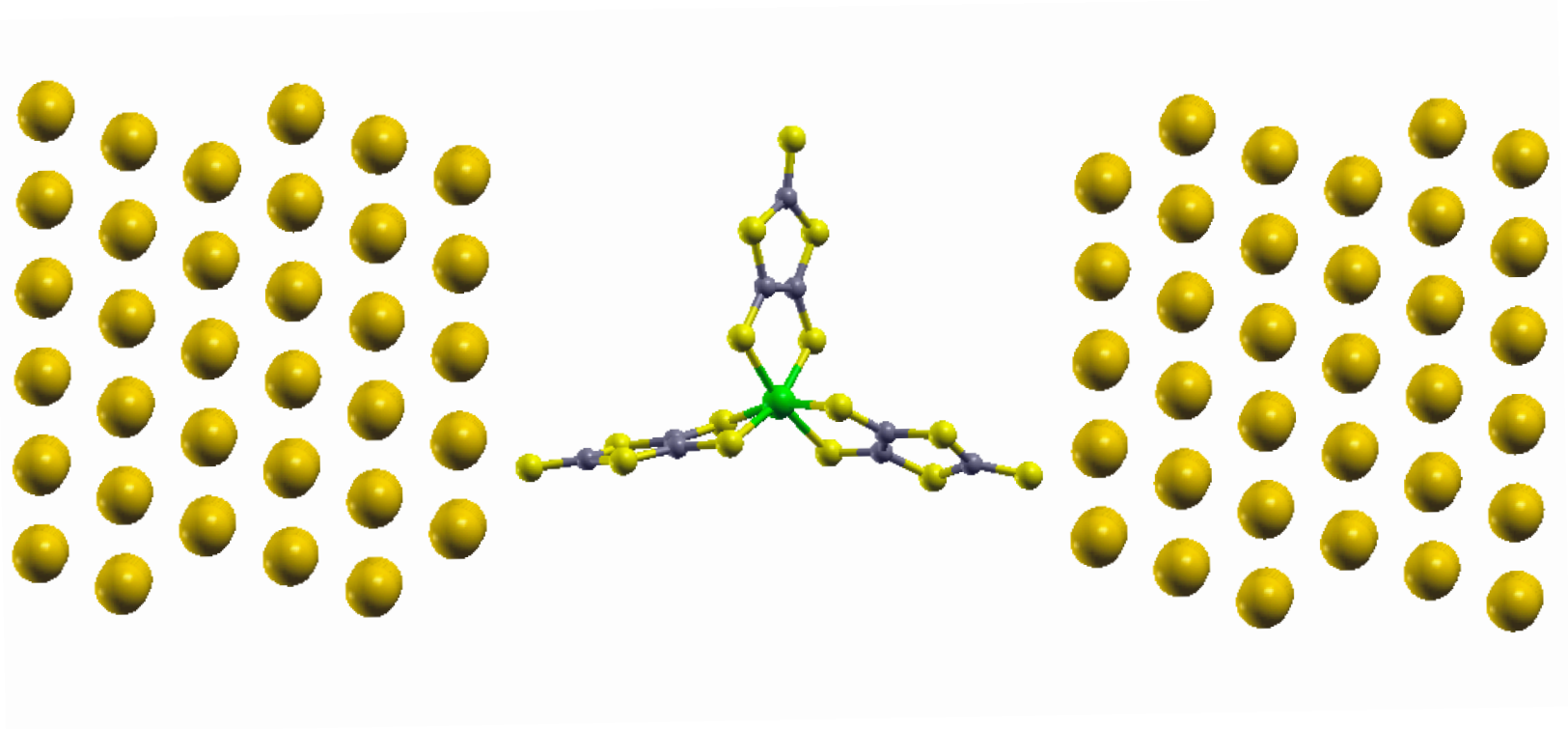

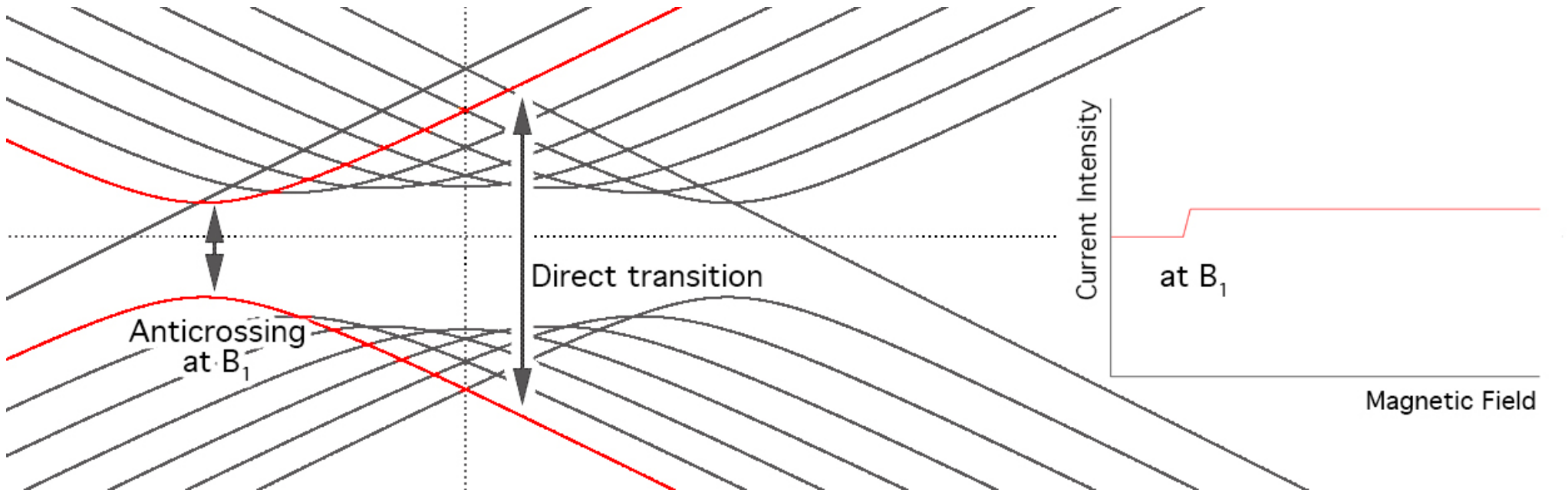

**Figure 1:** (Top panel) Calculated structure of $[V(\alpha\text{-}C_3S_5)_3]^{2-}$ after the structural relaxation (see Supplementary Information) while attached to two Au(111) electrodes. (Color code: Au = dark yellow, C = gray, S = yellow, V = green). (Bottom panel) Pictorial scheme of the hyperfine-controlled single-molecule spin transistor for a $I$ = 7/2 system, showing the anticrossing in the nuclear splitting, where (at $B_1$) an abrupt jump in the current intensity would be measured. Using a slowly relaxing electron spin minimizes the direct transitions and thus improves the signal/noise ratio.

In order to evaluate whether the apparent advantages of $[V(\alpha\text{-}C_3S_5)_3]^{2-}$ can be used in practice, one needs to verify two remaining requirements (a) the conductance needs to be spin-dependent at a reasonable gate voltage and, (b) for a chosen transmission peak, there must exist a conduction channel that is compatible with the survival of the electronuclear spin state. This second point is critical. In fact it is easy to imagine an organic radical or a metal complex where the electron spin resides at the Fermi level, $E_F$, and in that case the electron spin state would not be well defined in the conduction regime.

In this work, we have evaluated these two conditions by theoretically studying the transport properties of a two-terminal junction made of Au(111)-oriented electrodes sandwiching $[V(\alpha\text{-}C_3S_5)_3]^{2-}$. The theoretical transport calculations are based on the standard NEGF-DFT framework using the SMEAGOL[17] code, where atomic self-interaction correction (ASIC) is included in order to avoid the pinning of the HOMO/LUMO to $E_F$ (details about the procedure are shown in Supplementary Information).[18,19]

The first questions one has to answer are: does $[V(\alpha\text{-}C_3S_5)_3]^{2-}$ conduct current at all and, if so, is the current spin-dependent; furthermore is a gate voltage required for driving a spin-dependent current? Since the experiments are usually conducted for an infinitesimal source-drain voltage and at a low enough temperature to ensure that the electronic spin is fully blocked, we have calculated the zero-bias transmission spectra, and distinguished between spin-up and spin-down transmissions. We have limited our calculations to an energy range corresponding to the gate voltage range explored in the original experiments, namely up to ±2 V. In order to obtain an initial, rough estimate of the energy overlap between the conductance pathway and the magnetic orbitals, we have also computed the projected density of states (PDOS) on the vanadium ion. These two molecular fingerprints are shown superimposed (see Figure 2, upper panel).

The transmission plotted in Figure 2 manifests two overlapping conductance peaks with very strong spin asymmetry at and around $E_F$, as well as four sharp peaks at accessible gate voltages, which also present very strong spin asymmetry. At any of these peaks and in the presence of a magnetic field, the expected conductance behavior of the spin-up configuration of the molecule is very different from the expected behavior for the spin-down configuration. Therefore, we could expect to see the kind of jump in the conductance accompanying a flip in the electronuclear spin state, as experimentally observed at each anticrossing in the hyperfine-controlled single-molecule spin transistor experiment.[5] There is also a broad conductance peak 1eV below $E_F$, but for that peak the spin dependence is more complex and the spin asymmetry is weaker. Among all conductance peaks calculated for $[V(\alpha\text{-}C_3S_5)_3]^{2-}$, the vanadium-projected density of states is minimal at the Fermi level, thus we will be assuming no gate voltage is applied throughout the remainder of this work.

In order to check the properties of the junction at finite bias, we have calculated the current-voltage, *I-V*, characteristics at zero gate voltage, by varying the bias voltage both in the conventional range [-1.0:1.0] V and zooming in at small bias [-1.0:1.0] mV as done in the original experiments. In both cases, we have worked in the same low-temperature regime that was employed experimentally to ensure full polarization of the electronic spin state of the molecule: our calculations are therefore performed at an electronic temperature of $T = 0.1$ K. The calculated *I-V* curve (see Fig. 2, lower panels) shows that the spin current is almost entirely polarized for voltages between -1 and +1 V. The spin up component of the spin current is at least three orders of magnitude more intense than the spin down one, meaning that the transmission path is influenced by the molecular spin. The fact that the molecule conducts at zero gate voltage and infinitesimal bias voltage is due to a lucky arrangement of the hybridized molecular orbitals and the electrode work function; note that other analogous vanadium complexes where the same calculations were performed did not display this feature, (see SI and Ref. 20). One should notice that the large value found for the total molecular conductance is unsurprising, since we have looked here at a situation of strong chemical bonding between the molecule and the electrodes.

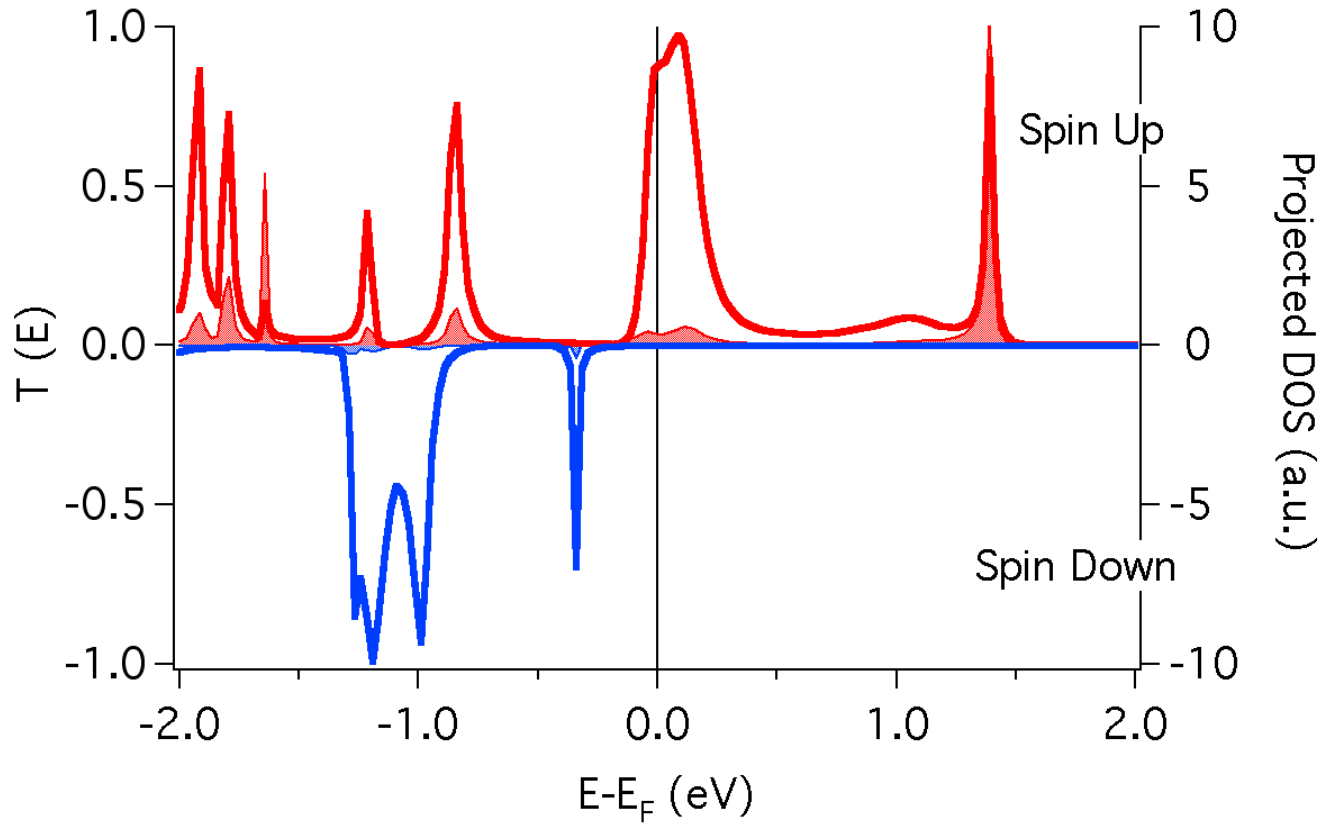

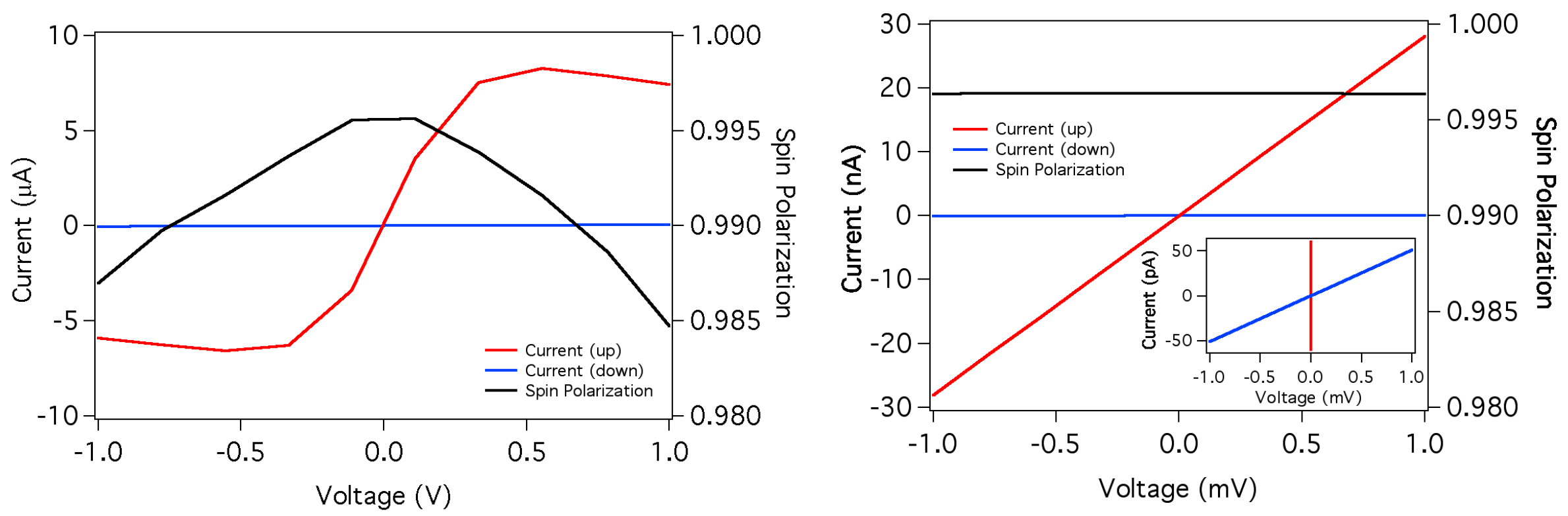

**Figure 2:** Upper panel: Spin dependent transmission spectra (left axis, solid lines) and vanadium-projected density of states (right axis, shaded area) of $[V(\alpha\text{-}C_3S_5)_3]^{2-}$. Note that the spin-down quantities are plotted on the negative axis for clarity. Lower panel: Current-voltage curves in two different voltage ranges. The red (blue) lines represent the spin-up (spin-down) current. Inset: vertical zoom for the same voltage range to visualize the extremely shallow slope of the spin down current. The black line represents the spin polarization ratio.

So far we have confirmed that $[V(\alpha\text{-}C_3S_5)_3]^{2-}$ presents a spin-dependent transport channel at zero gate and infinitesimal bias voltage, with little participation of the vanadium orbitals. By analyzing the local Density of States (LDOS) at the Fermi energy (Figure 3, upper panel) one can assert that the main conduction channel is formed by the *p*-type orbitals of sulfur and carbon along the horizontal ligand chain with a contribution of the *d*-type orbitals of the vanadium ion. Since the spin is expected to reside in the V *d* shell, it is necessary to evaluate the relation between the spin orbitals and the conduction channel, taking into account the spin delocalization over the entire molecule and the actual energy of the occupied spin orbitals. We have integrated the DOS up to $E_F$ and calculated its space-resolved distribution (partial density of states of Figure 3, lower panel). From the spin distribution we can infer that the majority of the unpaired electron density is located either on the vanadium ion or delocalized over the vertical $[\alpha\text{-}C_3S_5]^{2-}$ ligand.

A visual inspection indicates that occupied spin orbitals and conduction channel are orthogonal. At the Fermi level, a maximum electron density can be found in the sulfur atoms that bond to the vanadium ion and the electron density at the upper $[\alpha\text{-}C_3S_5]^{2-}$ is cut through a clean node. In contrast, the space-resolved electron density with residual spin is practically zero for the sulfur atoms that bond to the vanadium ion and no node is present in the upper $[\alpha\text{-}C_3S_5]^{2-}$ ligand. Furthermore, the Fermi vs residual spin densities that are present in the two lower $[\alpha\text{-}C_3S_5]^{2-}$ are also orthogonal. From a quantitative point of view and focusing on the vanadium ion, we analyzed the overlap between its magnetic orbitals and the conduction channel (see Fig 3, lower panel). In such analysis we found that the net spin residing on the V atom, presents a low contribution at $E_F$, indicating that one could have an effective current along the molecule without directly affecting the unpaired spin, since the net spin density resides mostly in orbitals that are at least 3 eV below $E_F$,. This theoretical estimate of a redox-robust spin coincides with previous experimental electrochemistry experiments where the vanadium oxidation state is maintained during the single and double oxidation of the complex.[21] We have thus checked the second main condition: the electronuclear spin states must be able to survive in the conduction regime. Moreover, the fact that there is virtually no spin density on the vanadium *s* orbitals helps stabilize the nuclear spin qubits (see Supplementary Information).

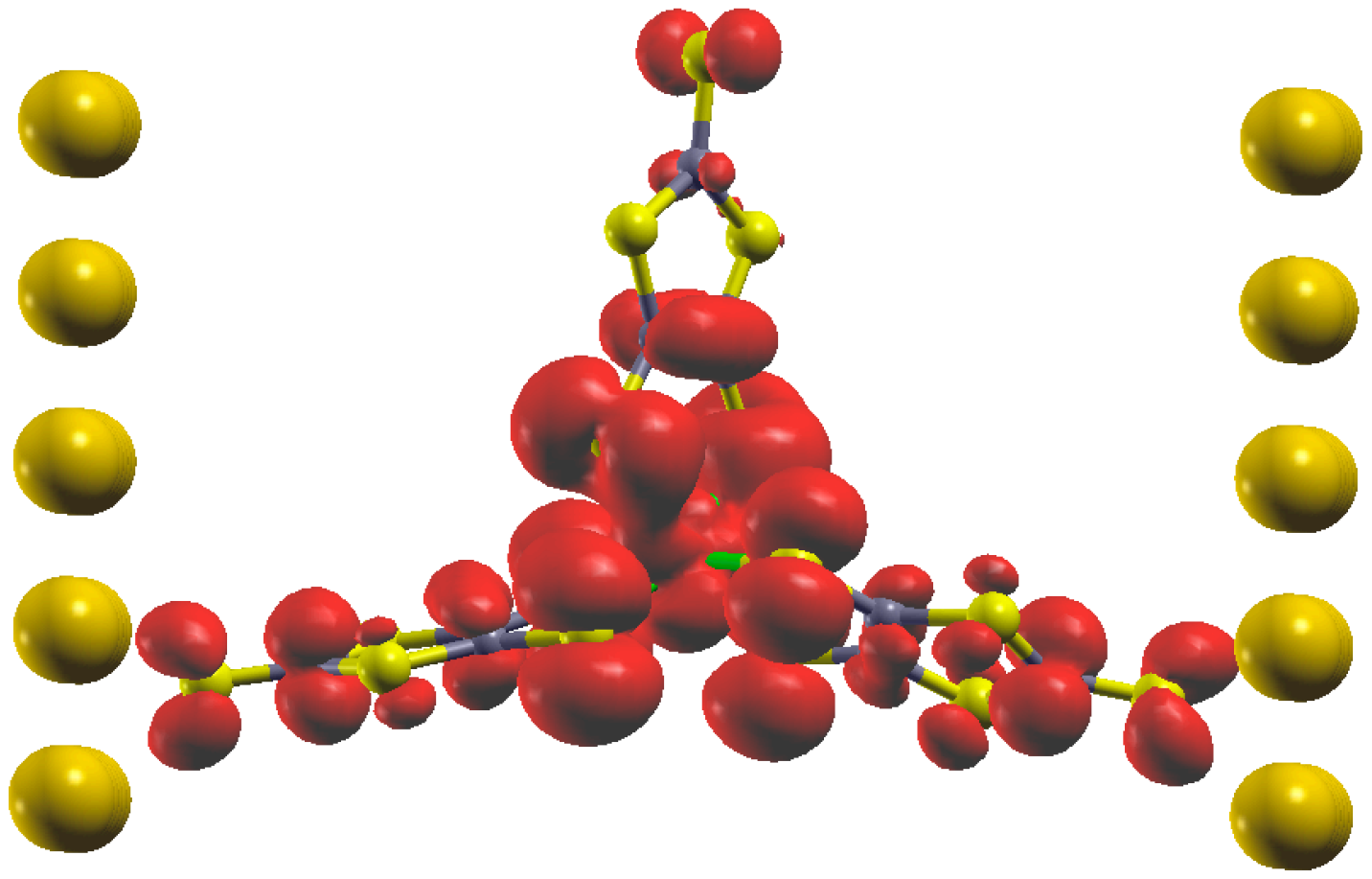

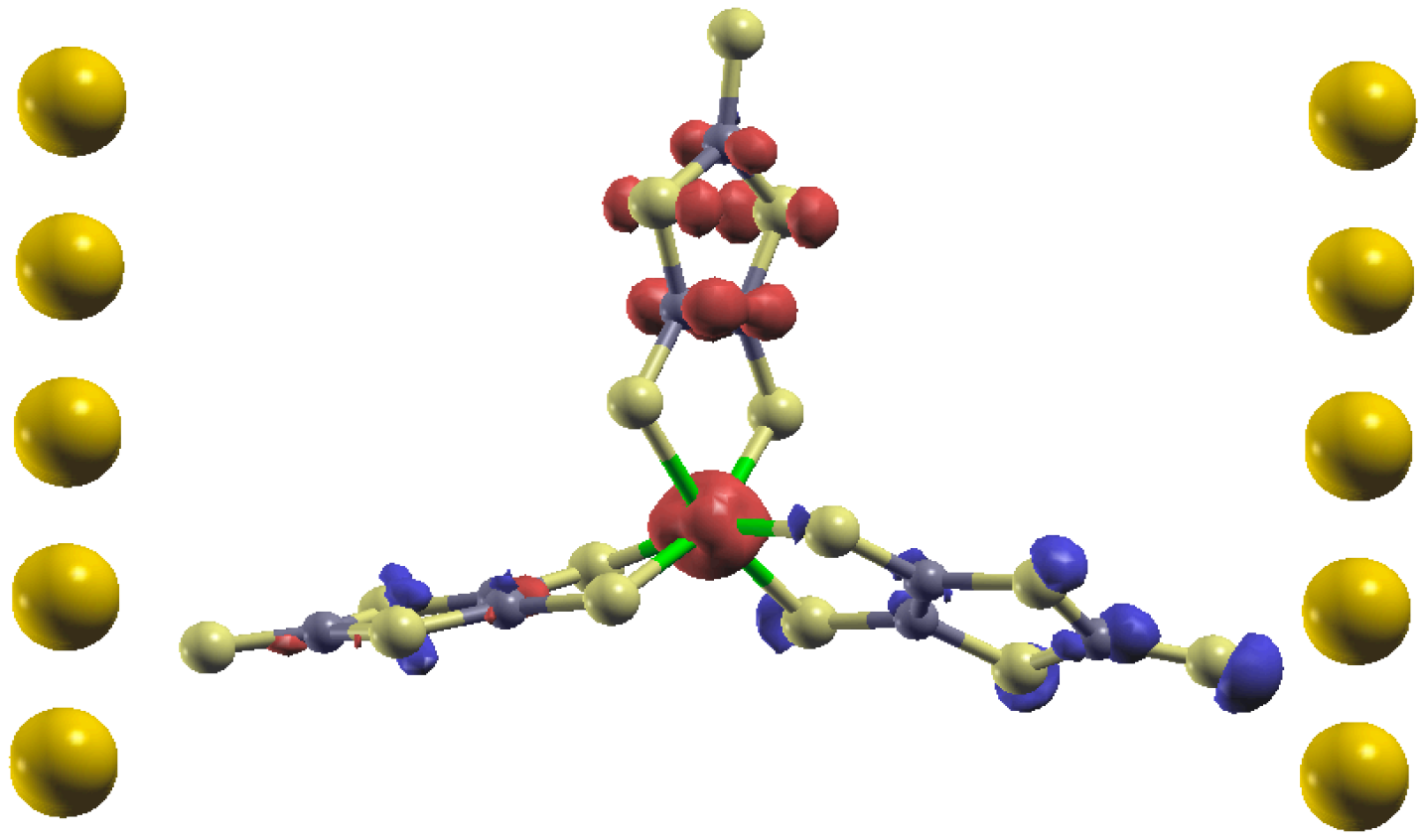

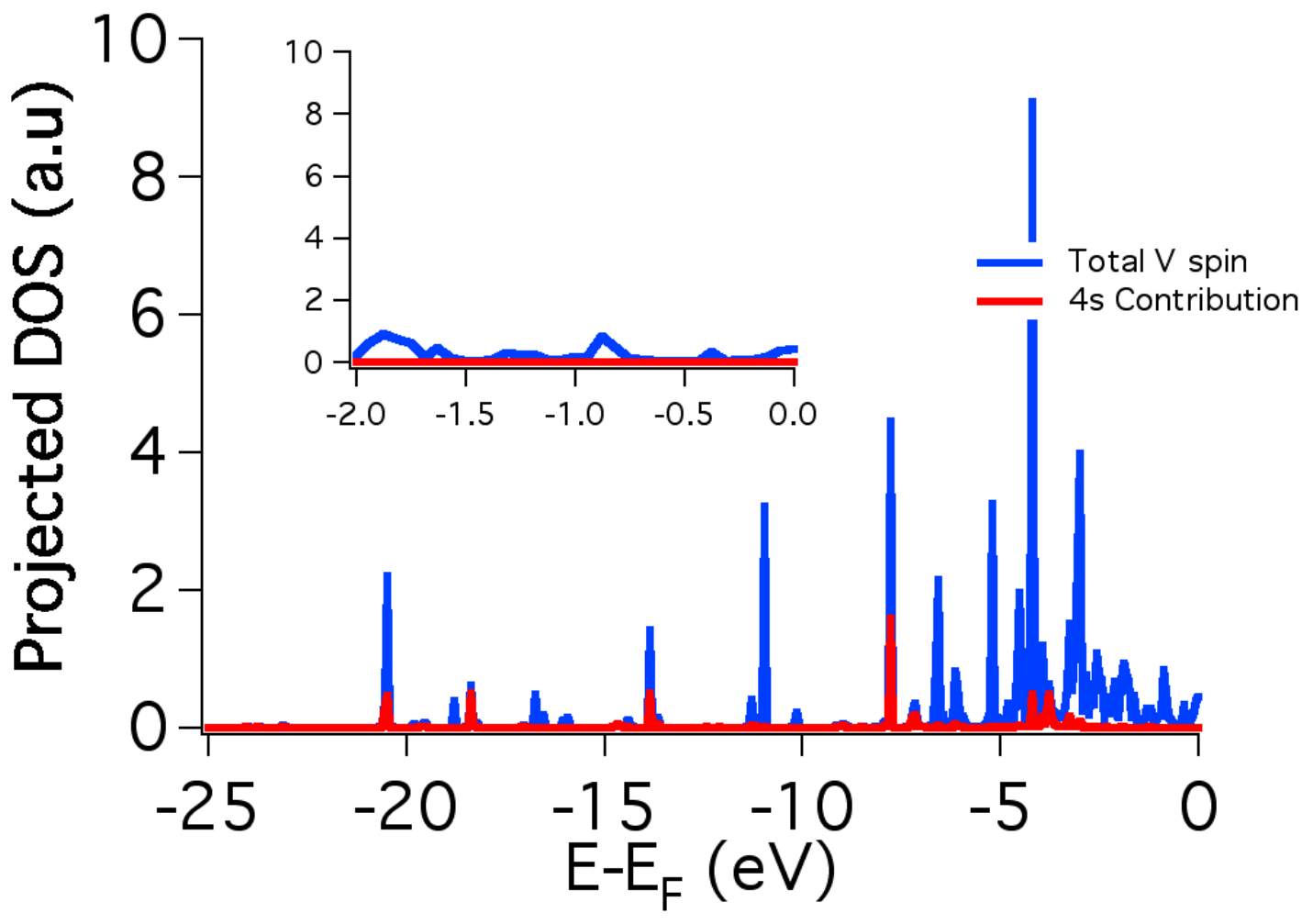

**Figure 3:** Upper panel: Local density of states at Fermi level. Mid panel: space-resolved electron density for $[V(\alpha\text{-}C_3S_5)_3]^{2-}$, this is the DOS integrated from -25 eV up to the Fermi level. Red (blue) density represents spin up (down) electronic density. Lower panel: Total net spin (spin up-spin down) for the vanadium and its projection on 4s orbital $[V(\alpha\text{-}C_3S_5)_3]^{2-}$.

In summary, we have explored here the complex $[V(\alpha\text{-}C_3S_5)_3]^{2-}$, and found it presents certain advantages when compared with the system employed so far, $[Tb(Pc)_2]^-$. Being a vanadium complex, it can host three nuclear spin qubits instead of two, and it presents a tunneling splitting that is at least three orders of magnitude above $1\mu$K, the value found for $[Tb(Pc)_2]^-$, which would allow one to sweep the magnetic field as fast as desired, alleviating an experimental limitation. More specifically, $[V(\alpha\text{-}C_3S_5)_3]^{2-}$ guarantees a good attachment to the electrode through the well-studied Au-S hybridization, and moreover it presents a good environment for the nuclear spin qubit, since the relaxation of the electron spin is known to be slow in this complex which contains no further nuclear spins. Crucially, our calculations reveal that in this system (a) the conduction pathway has a minimal overlap with the orbital hosting the electron spin and (b) it should be possible to obtain a conductance that is strongly asymmetric with respect to the molecular spin, with no need for a gate voltage and applying a minimal bias voltage. The only design criterion which is not met by $[V(\alpha\text{-}C_3S_5)_3]^{2-}$ is that, as far as the hyperfine parameters determined in the bulk can be used for a single molecule attached to gold, it does not satisfy $A_\parallel > A_\perp$ and therefore it could present an inconveniently small separation between anticrossings in terms of the applied magnetic field. It is now evident that it is both urgent and feasible to explore further candidate molecules for the electric readout and manipulation of nuclear spins.

AUTHOR INFORMATION

**Notes**

The authors declare no competing financial interests.

ACKNOWLEDGMENT

The research reported here was supported by the Spanish MINECO (Grants MAT 2014-56143-R and CTQ 2014-52758-P co-financed by FEDER, and Excellence Unit María de Maeztu MDM-2015-0538), the European Union (ERC-CoG DECRESIM 647301 and COST 15128 Molecular Spintronics Project), and the Generalitat Valenciana (Prometeo Program of Excellence). A. Gaita-Ariño thanks the Spanish MINECO for a Ramón y Cajal Fellowship. S. Cardona-Serra acknowledges the Generalitat Valenciana for a VAL i+D postdoctoral contract. M. Stamenova acknowledges support by Science Foundation Ireland (grant No. 14/IA/2624). S. Sanvito acknowledges the Quest project funded by European Research Council. We thank the Trinity Centre for High Performance Computing (TCHPC) for providing the computational resources.

[1] S. Sanvito*, Chem. Soc. Rev.*, 2011, **40**, 3336.

[2] V. A. Dediu, L. E. Hueso, I. Bergenti, C. Taliani, *Nat. Mater.* 2009, **8**, 707.

[3] L. Bogani, W. Wernsdorfer, *Nat. Mater.* 2008, **7**, 179.

[4] S. Thiele, F. Balestro, R. Ballou, S. Klyatskaya, M. Ruben, W. Wernsdorfer, *Science* 2014, **344**, 1135-1138.

[5] S. Thiele, PhD Thesis 2016. Springer PhD Theses series. ISBN: 978-3-319-24058-9

[6] W. Wernsdorfer, APS Meeting 2016, http://meetings.aps.org/link/BAPS.2016.MAR.F21.1.

[7] AlDamen, M.; Cardona-Serra, S.; Clemente Juan, J. M.; Coronado, E.; Gaita-Ariño, A.; Martí-Gastaldo, C.; Luis, F.; Montero, O. *Inorg. Chem* **2009**, *48* (8), 3467.

[8] Shiddiq, M.; Komijani, D.; Duan, Y.; Gaita-Ariño, A.; Coronado, E.; Hill, S. *Nature* **2016**, *531* (7594), 348.

[9] Jenkins, M. D.; Duan, Y.; Diosdado, B.; García-Ripoll, J. J.; Gaita-Ariño, A.; Giménez-Saiz, C.; Alonso, P. J.; Coronado, E.; Luis, F. *Phys. Rev. B* **2017**, *95* (6), 064423.

[10] Cardona-Serra, S.; Clemente Juan, J. M.; Coronado, E.; Gaita-Ariño, A.; Suaud, N.; Svoboda, O.; Bastardis, R.; Guihéry, N.; Palacios, J. J. *Chem-Eur J* **2014**, *21* (2), 763.

[11] Lehmann, J.; Gaita-Ariño, A.; Coronado, E.; Loss, D. *Nature Nanotechnology* **2007**, *2* (5), 312.

[12] Sherif, S.; Rubio-Bollinger, G.; Pinilla-Cienfuegos, E.; Coronado, E.; Cuevas, J. C.; Agrait, N. *Nanotechnology* **2015**, *26* (29), 291001.

[13] Zadrozny, J. M.; Niklas, J.; Poluektov, O. G.; Freedman, D. E. *ACS Central Science* **2015**, *9*, 488.

[14] Yu, C.-J.; Graham, M. J.; Zadrozny, J. M.; Niklas, J.; Krzyaniak, M. D.; Wasielewski, M. R.; Poluektov, O. G.; Freedman, D. E. *Journal of the American Chemical Society* **2016**, *138* (44), 14678.

[15] Bader, K.; Winkler, M.; Van Slageren, J. *Chem. Commun.* **2016**, *52*, 3623.

[16] Atzori, M.; Tesi, L.; Morra, E.; Chiesa, M.; Sorace, L.; Sessoli, R. *Journal of the American Chemical Society* **2016**, *138* (7), 2154.

[17] A. R. Rocha, V. Garcia-Suarez, S. W. Bailey, C. J. Lambert, J. Ferrer, S. Sanvito, *Nat. Mater.* **2005**, *4*, 335.

[18] C. Toher and S. Sanvito, *Phys. Rev. Lett.*, **2007**, *99*, 056801.

[19] C. D. Pemmaraju, T. Archer, D. Sánchez-Portal and S. Sanvito, *Phys. Rev. B*, **2007**, *75*, 045101– 045116.

[20] Cardona-Serra, S. et al. Work in preparation.

[21] Akiba, K.; Matsubayashi, G. E.; Tanaka, T. *Inorganica Chimica Acta* **1989**, *165* (2), 245.

# Supporting Information for:

## Theoretical evaluation of $[V((\alpha\text{-}C_3S_5)_3]^{2-}$ as nuclear-spin sensitive single-molecule spin transistor.

*S. Cardona-Serra*[1,2], A. Gaita-Ariño[1], M. Stamenova[2] and S. Sanvito[2].*

[1] Instituto de Ciencia Molecular, Universitat de València, España.

[2] School of Physics, CRANN, Trinity College Dublin 2, Ireland.

Contents:

1: Theoretical Methodology

First-principles calculations are performed using the SMEAGOL code[1] that interfaces the non-equilibrium Green's function (NEGF) approach to electron transport with the density functional theory (DFT) package SIESTA[2].

In our simulations the transport junction is constructed by placing the molecule between two Au (111)-oriented surfaces with 5×5 cross section. This mimics a standard transport break-junction experiment with the most used gold surface orientation. The choice of a gold electrode arises from the stability of the sulfur-gold bond that ensures the best attachment between the molecule and the surfaces. The initial S-surface distance was set to 2.0 Å with the S atom located at the 'hollow site', the most stable absorption position discussed in literature.[3] Thus, the entire structure is then relaxed until the maximum atomic forces are less than 0.01 eV/Å. A real space grid with an equivalent plane wave cutoff of 200 Ry (enough to assure convergence) has been used to calculate the various matrix elements. Finally the electronic temperature of the calculation (unless specified) is set to 0.1 K to mimic the low-temperature limit conditions used in the original experiment. During the calculation the total system is divided in three parts: a left-hand side lead, a central scattering region (SR) and a right-hand side lead. The scattering region contains the molecule as well as 4 atomic layers of each lead, which are necessary to relax the electrostatic potential to the bulk level of Au.

The convergence of the electronic structure of the leads is achieved with 2x2x128 Monkhorst-Pack k-point mesh, while for the SR one sets open boundary conditions in the transport direction and periodic ones along the transverse plane, for which an identical k-point mesh is used (2x2x1 k-points). The exchange-correlation potential is described by the Ceperley-Alder local density approximation (LDA)[4] as implemented for using the ASIC approach (see below). One of the main problems using local/semilocal DFT functionals for the study of the interfaces between inorganic leads and molecules is the self-interaction error. This leads to an over-delocalization of the electronic density producing the HOMO orbitals to be set higher in energy artificially. The LUMO orbitals could also be found at lower energy values that the commonly expected. In order to avoid this problem we have used the Atomic Self Interaction Correction (ASIC) method.[5] This methodology is known to correctly set the position of the HOMO orbital, and it has also been shown to improve the energy level alignment when the junction is formed, leading to values of conductance in better agreement with experiments.[6]

The Au-valence electrons are represented over a numerical s-only single-ζ basis set that has been previously demonstrated to offer a good description of the energy region around the Fermi level.[7] In contrast, for the other atoms (S, C, V and O) we use a full-valence double-ζ basis set with polarization (basis size was increased until convergence). Norm-conserving Troullier-Martins pseudopotentials[8] (adapted to ASIC methodology) are employed to describe the core-electrons in all the cases. Considering the fact that all the molecules studied here are dianions, we have shifted the energy levels obtained by adding (removing) a specific number of charges depending of the molecular charge. The limitation of this approach is that these charges are only added for the calculation of the electrostatic potential but do not enter in the self-consistent charge density, thus they will only interact indirectly with the charges of the scattering region via the electrostatic potential. (See SMEAGOL manual for more details about calculations on charged systems).

Finally, the spin-dependent current, $I_\sigma$, flowing through the junction is calculated from the Landauer-Büttiker formula[9],

$$I_\sigma(V) = \frac{e}{h}\int_{-\infty}^{+\infty} T_\sigma(E,V)[f_L(E-\mu_L) - f_R(E-\mu_R)]dE$$

where the total current $I_{tot}$ is the sum of both the spin-polarized components , $I_s$, $\sigma$ = spin up /spin down. Here $T_\sigma(E,V)$ is the transmission coefficient[1] and $f_{L/R}$ are the Fermi functions associated to the two electrodes chemical potentials, $\mu_{L/R} = \mu_o \pm V/2$, where $\mu_o$ is the electrodes common Fermi level.

In our two-spin-fluid approximation (there is no spin-flip mechanism) majority and minority spins carry two separate spin currents and the resultant current spin polarization, *SP*, is calculated as

$$SP = \frac{I_{up} - I_{down}}{I_{up} + I_{down}}$$

Note that although this approach is widely used to perform theoretical predictions within the spintronic community,[10] it has some limitations inherent to the basics of DFT methodology as pointed out by Lozano et al.[11] One may note that the theoretical spin asymmetry corresponds to what is expected from an open-shell DFT calculation in a magnetic molecule. Thus, within the mono-determinant approximation of DFT only a concrete $M_S$ state is calculated. In literature this has been sometimes confused with the obtainment of a super-efficient spin filter. This is a different scenario compared with our case, where the low temperature guarantees the spin orientation is experimentally fixed by an external magnetic field.

2: Detailed comments on the selection of molecules for the electrical readout of nuclear spin

A wide variety of chemical systems can in principle be candidates for this kind of experiment. Taking a magnetic field and conducting molecule with hyperfine levels, the main requisite would be that both the electronic and the nuclear spin states survive in a conduction regime. In practice, this means that the molecular orbitals hosting the electron spins should not be directly part of the conducting pathway; this protects the electron spin states and often guarantees implicitly that the contact hyperfine coupling between the nuclear spin and the conducting pathway is negligible, thus protecting the nuclear spin states. Additionally, of course, long spin-lattice relaxation times are desirable since they reduce operating noise and increase the maximum operating time. Furthermore, a large quantum tunneling also reduces operating noise and enhances operating speed.

The first main group to consider would be lanthanide-based single ion magnets, such as $TbPc_2$ used by Wernsdorfer et al. This family of complexes has grown to several hundred, and there is a continuous effort to develop strategies to control their magnetic behavior.[12] A problem, of course, is that the majority of the complexes will be electrical insulators in their most chemically stable charge states, but there are a number of exceptions. Notably, beyond the phthalocyaninato sandwiches, the obvious candidates would be polyoxometalate complexes,[13,14] which are known for their ability to host a varying number of electrons without suffering structural alterations;[15,16] further possibilities, with the possible drawback of the smaller size, would be organometallic aromatic complexes.[17] Among polyoxometalates, a most promising family would be the single ion magnets and molecular spin qubits $LnW_{30}$ [18,19] and in particular $TbW_{30}$ which has experimentally be proven to function as a single-molecule rectifier.[20] We chose not to start our theoretical study with these promising families purely because of reasons of computational cost: Lanthanoid complexes need an explicit consideration of the spin-orbit coupling with relativistic pseudopotentials that would make the initial exploration much more challenging. In addition, except for some organometallic 'sandwiches', all the lanthanoid single ion magnets present a larger number of atoms per molecule compared with other simpler families.

A second chemical family with potential interest are the recent and smaller family of transition-metal based molecular spin qubits.[21,22,23,24] In this case an obvious advantage would be the long relaxation times, if one can assume that the times will be maintained in a vastly different environment. Again, electrical conductance would need to be verified for these systems, but the complexes are promising since they tend to present planar structures stabilized by ample aromatic π clouds. The critical problem here is the survival of the spin state: since *d* orbitals participate in bonding, there is in principle a high risk that the spin resides in the same orbitals that participate in the conduction process.

To choose a starting candidate from the family of transition metal ions we employed a series of criteria. First, again for reasons of computational cost, we discard the phthalocyaninato complexes.[24] Next we consider the practical aspect of attaching the molecule to electrodes and, at the same time, the computational resources that are linked to the magnitude of the nuclear spin. From both aspects, we discard the $Cu^{2+}$ complexes (I=3/2, 2 qubits)[23] and choose the $V^{4+}$ complexes (I=7/2, 3 qubits), and among them those where the ends of the molecule are sulfur atoms which can be attached to gold.[21,22] These molecules are (bis-) tris-dithiolene metal complexes that although being known for a long time have been revisited using new techniques, modern instrumentation and advanced computational methodologies.[25] Finally, among those, there are advantages and disadvantages whether one chooses a vanadium complex or a vanadyl one. The advantage of vanadyl is its hyperfine coupling, which -as far as one can trust it to be unperturbed by the hybridized molecular orbitals after attachment to gold- is far more promising, since it allows an easy distinction between the different transitions (see Section 6: Extended discussion on hyperfine structure and anticrossings). However, preliminary calculations showed that the vanadium complexes in ref. 21 presented a significant number of spin-selective transmission pathways while the conductance for the vanadyl complexes in ref. 22 was spin-symmetrical. See for example, transmission spectra of $[VO(\alpha\text{-}C_3S_5)_2]^{2-}$, Figure S.0 [26]

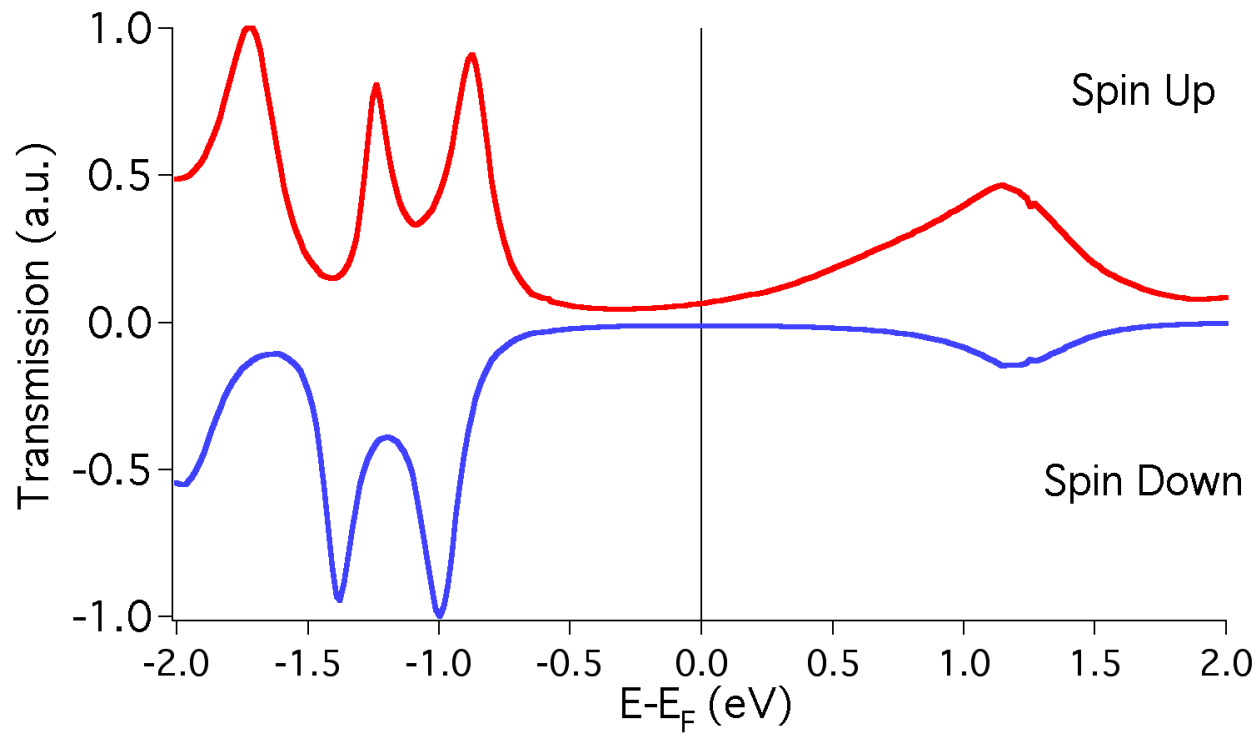

**Figure S0:** Spin dependent transmission spectra $[VO(\alpha\text{-}C_3S_5)_2]^{2-}$. Note that the spin-down quantity is plotted on the negative axis for clarity.

3: Extension of the present work to two analogous compounds

To verify the robustness of the calculations presented in the main text, we extended the same methodology to two analogous compounds, depicted in Figure S1.

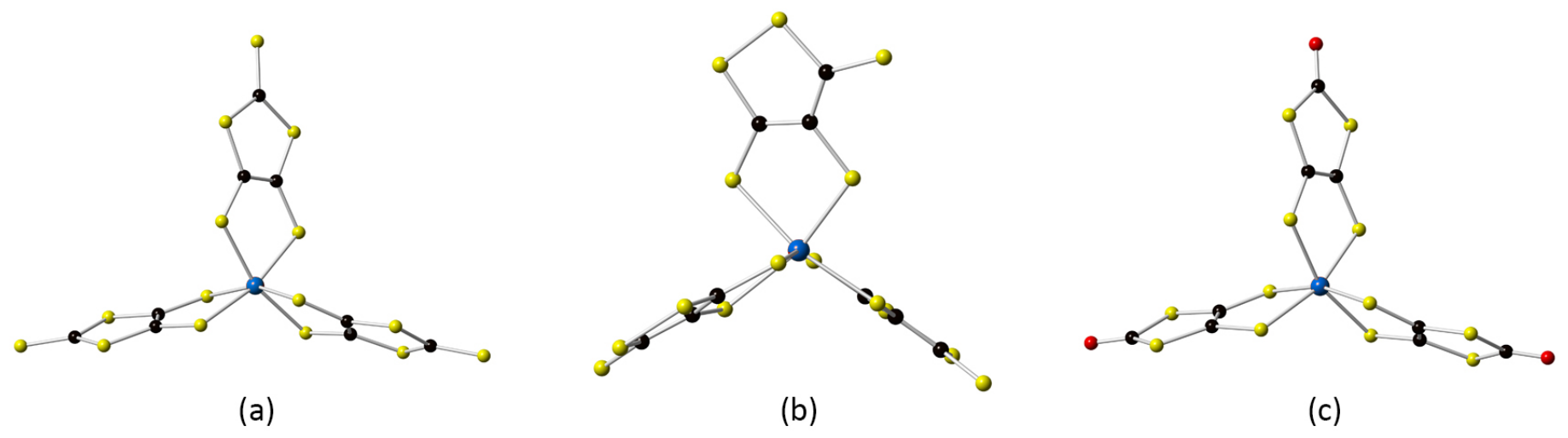

**Figure S1:** Molecular structure of the molecule presented in the main manuscript: (a) $[V(\alpha\text{-}C_3S_5)_3]^{2-}$ **1**, and two structural variations presented briefly as Supplementary Information: (b) $[V(\beta\text{-}C_3S_5)_3]^{2-}$ **2** and (c) $[V(\alpha\text{-}C_3S_4O)_3]^{2-}$ **3** (Colour code: Yellow: S, Black: C, Red: O, Blue: V) All three have been previously studied as highly stable molecular spin qubits.[21] The main difference between **1** and **2** is the S-S bond in the ligand and the different orientation of the terminal S atom which affects the attachment to the Au(111) electrodes. The main difference between **1** and **3** is a substitution of O for S in the terminal position, which is expected to decrease the hybridization with the Au(111) electrode.

As can be seen in Figure S2, in the case of the complex **2** the position of the various frontier orbitals are as follows: spin-up HOMO -0.21 eV, spin-up LUMO 1.49 eV; spin-down HOMO -0.31 eV and spin-down LUMO: 1.47 eV. In this case there's no available transport pathway with energy $E=E_F$. Then at infinitesimal bias there is not a clear contribution to the electronic transport and thus we expect the electronic current to be very low. We will demonstrate this consequence with the calculation of the *I-V*, Figure S3.

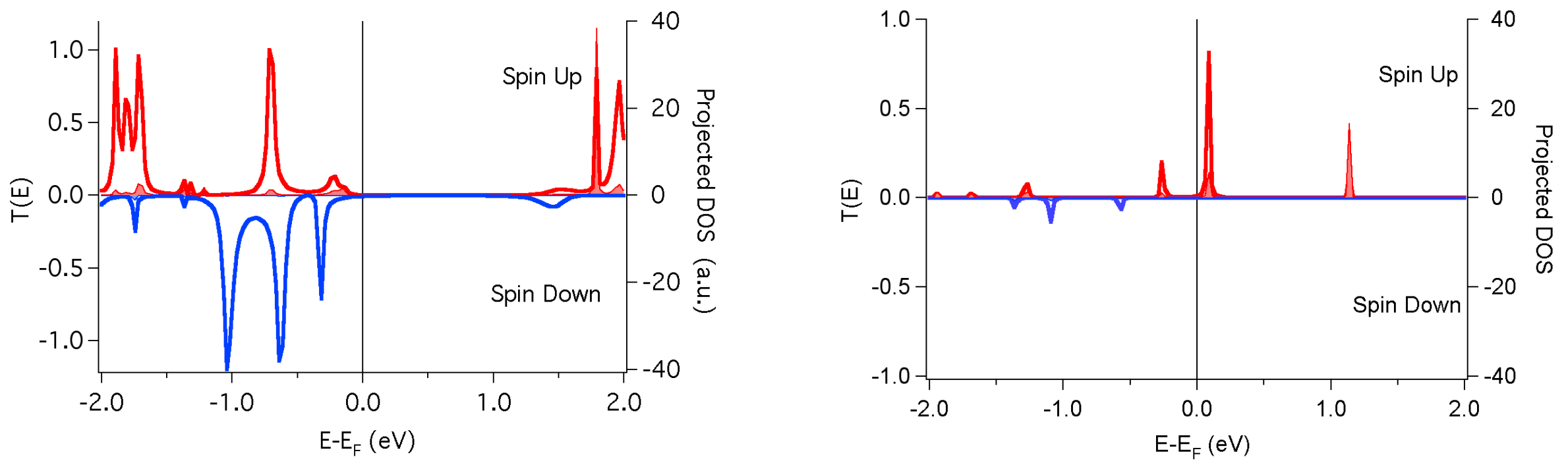

**Figure S2:** Spin dependent transmission spectra (Solid lines) and vanadium-projected density of states (shaded area) of $[V(\beta\text{-}C_3S_5)_3]^{2-}$ (**2**, left) and $[V(\alpha\text{-}C_3S_4O)_3]^{2-}$ (**3**, right). Note that the spin-down quantities are plotted on the negative axis for clarity.

Complex **3** presents the following energies: spin-up HOMO -0.26 eV; spin-up LUMO 0.09 eV; spin-down HOMO -0.56 eV; spin-down LUMO 3.07 eV. However, relating the width of the transmission peaks of the junction with the coupling between the molecule and the inorganic lead, it is clear that this is much smaller than that of the spectra of **1** and **2**, indicating that the hybridization between molecule and surface in **3** is much smaller than that in **1** and **2**. This results in a significantly lower current for the same bias range. We attribute this behavior to the fact that in **3** the terminal sulfur is replaced by an oxygen atom yet remaining in the same position than in **1**. As oxygen is harder than sulfur (in terms of Pearson classification) it is less capable of establishing a good orbital hybridization with gold.

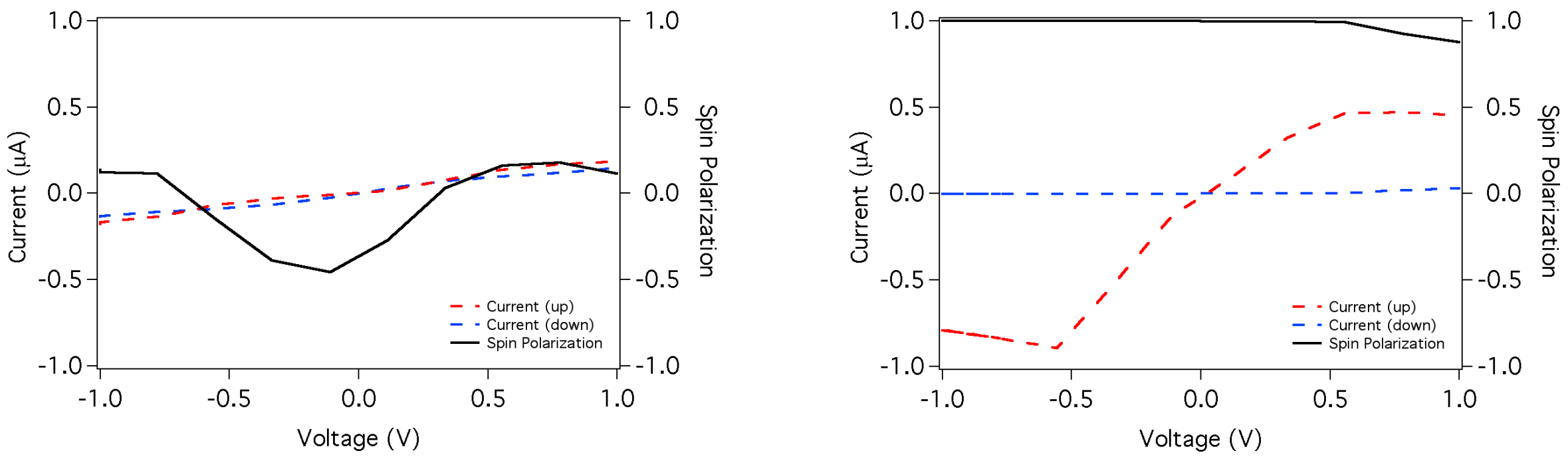

**Figure S3:** Current-voltage plot for the complexes $[V(\beta\text{-}C_3S_5)_3]^{2-}$ (**2**, left) and $[V(\alpha\text{-}C_3S_4O)_3]^{2-}$ (**3**, right). Red (Blue) dashed lines represent spin up (down) current. Solid black line represents the spin polarization ratio.

The *I-V* characteristic of complex **2** and **3** show a qualitative difference from that of the main complex. These two particular cases yield a current intensity that is up to one order of magnitude lower than such of the molecule **1**.

Analyzing the system **2** we particularize that we lack of any transmission pathway along the whole voltage window. In order to rationalize the voltage dependence, we have plotted the evolution of such transmission peaks with the increase of the bias voltage both positive and negative. (See figure S4, left). Although one could see a variable spin polarization along the voltage range, we have determined that such value is produced by the error noise intrinsically to the calculation method. The main chemical differences between this beta and the alpha isomers are:

1) the position of the apical sulfur atom that connects the molecule with the Au lead, and 2) the distance between the gold surface and the vanadium ion. In particular in **1** the S atom forms a line with the π-orbital of the ligand, while in **2**, it forms an angle of approximately 105°. Such difference in bonding geometry along with the shorter distance between the lead and the magnetic vanadium atom are the two main structural properties that affect the transmission paths in the two complexes.

In the complex **3** the replacement of the sulfur atom by oxygen is the main contributor to the differences observed in the I-V plot. This substitution partially impedes the electron flux through the junction, because of the lower hybridization between the molecule and the lead, thus producing a current that is significantly lower. The asymmetry obtained in the I-V plot maybe assigned to the relative uncertainty on the oxygen position due to the weaker oxygen-gold bond. In addition it is easy to see that the structural similarity of **1** and **3** manifests itself in the apparent resemblance in the intensity-bias voltage dependence.

4: Transmission spectra evolution with the applied bias voltage for **1**, **2** and **3**

Considering the previous I-V plots, we would like to analyze the evolution of the transmission spectra plot with the applied voltage. The displacement of the states peaks is crucial information to understand the spin polarization of the current.

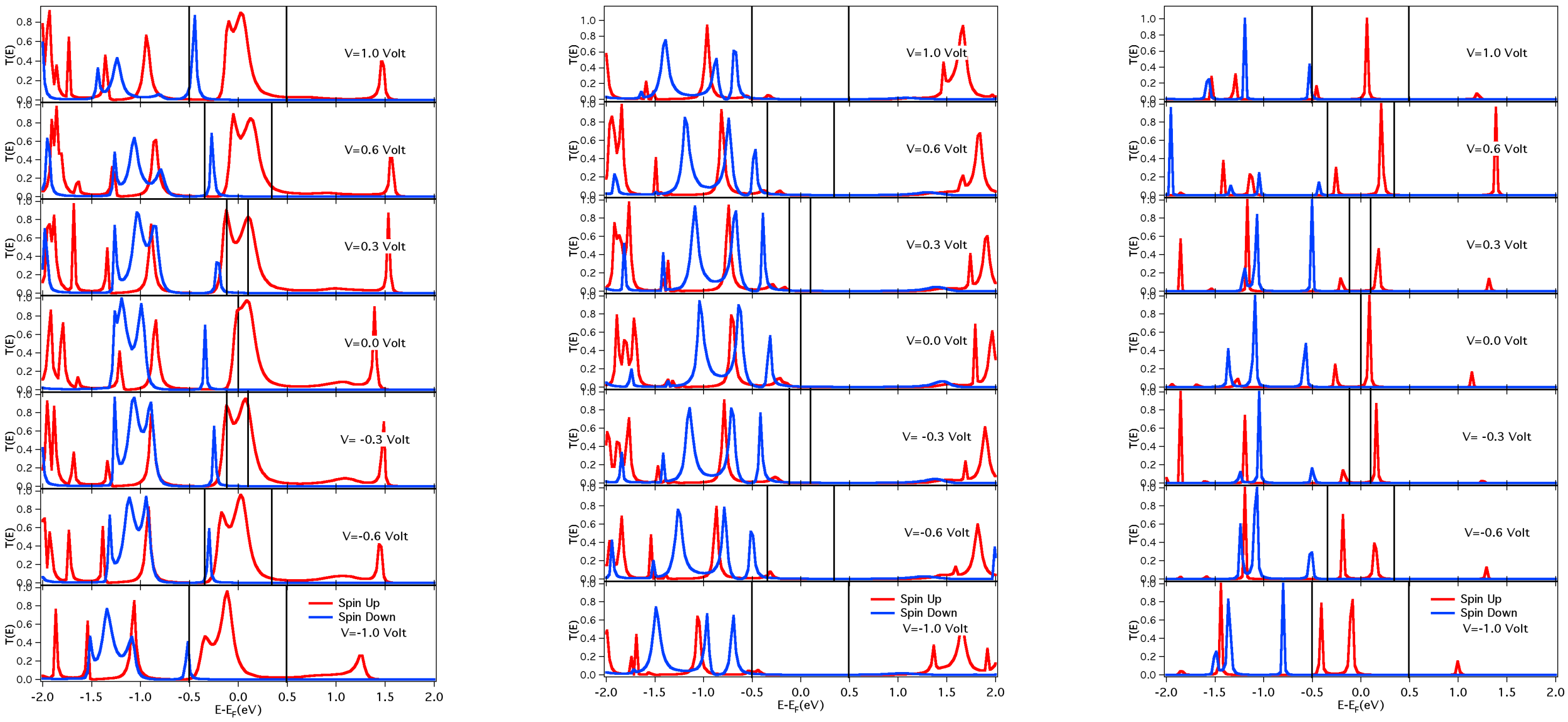

**Figure S4:** Evolution of the transmission spectra as a function of the applied voltage for the complexes $V(\alpha\text{-}C_3S_5)_3]^{2-}$ (**1**, left) $[V(\beta\text{-}C_3S_5)_3]^{2-}$ (**2**, mid) and $[V(\alpha\text{-}C_3S_4O)_3]^{2-}$ (**3**, right). Black lines represent the bias window.

5: Analysis of the spin projections in orbitals that participate in the conduction pathway

From a visual inspection it is clear that the four atoms where there is a possible overlap between the electron spin and the conduction pathway are the three carbon atoms in the upper ligand and the central vanadium atom. It is thus necessary to study their PDOS. In Figure S5 this is done for the three carbon atoms. As can be seen, the vast majority of the net spin in these atoms resides far away from the Fermi energy, and virtually nothing is available at the experimentally applied voltages: note that the horizontal scale in these graphs is [0,-25eV] ([0,-2eV] in the inset), while experiments are performed at a bias voltage around or below 1mV, so only the tiny peak centered at zero would participate.

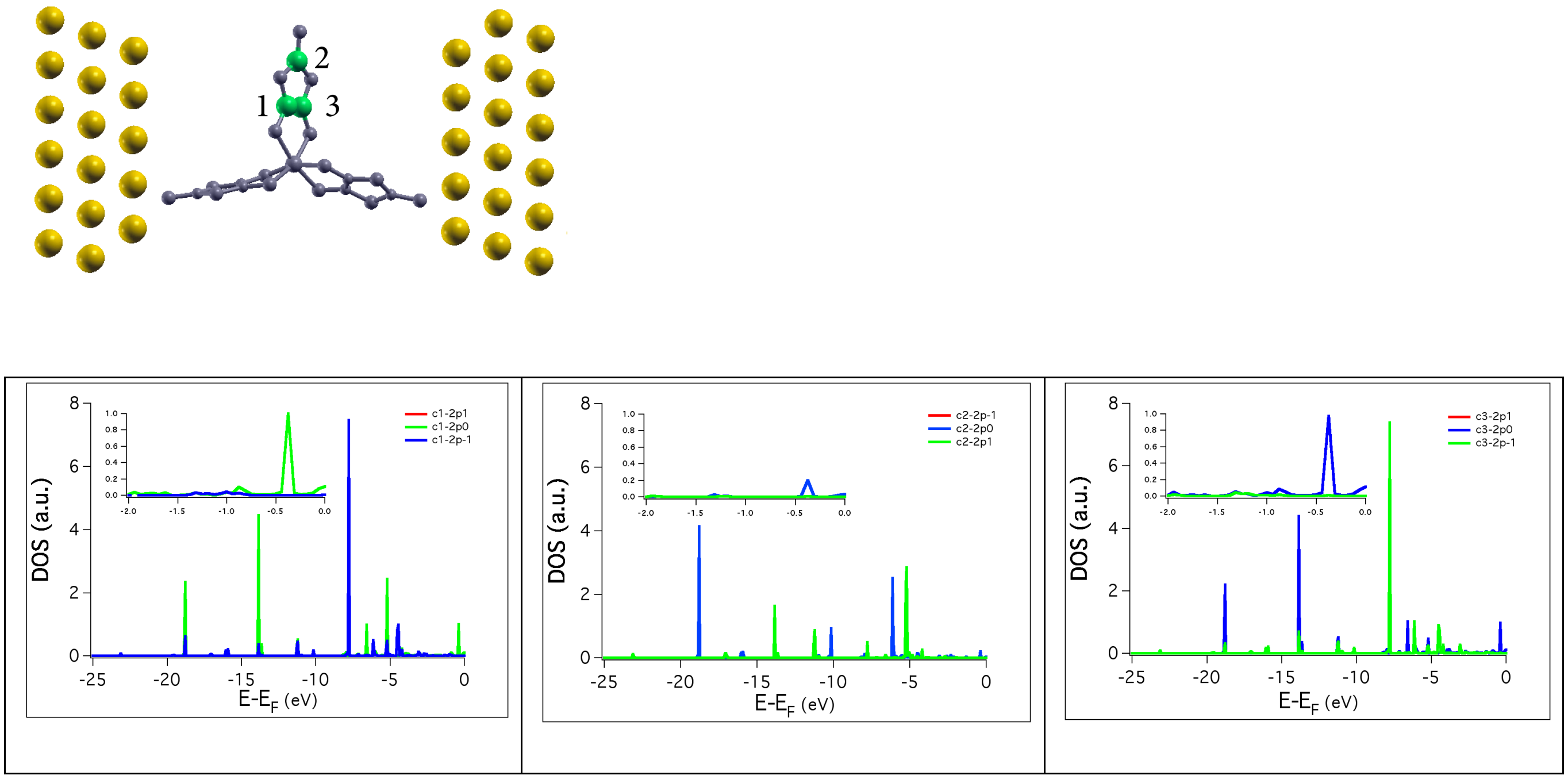

**Figure S5**: Net spin (spin up- spin down) projection on the vertical ligand carbon 2p orbitals for $[V(\alpha\text{-}C_3S_5)_3]^{2-}$ .

A very similar picture results from the analysis of the vanadium PDOS (Figure 3, lower panel). All in all, only a tiny fraction of the molecular spin is available to the conducting path, both in terms of energy and in terms of spatial distribution. The case of vanadium deserves a further study, since there would be a risk of direct flip of the nuclear spin state by the current if there were any detectable participation of the vanadium 4s orbitals in the Fermi level (or, in general, in the conduction pathway). Since only s-type orbitals are free from angular nodes, these are the only ones that are expected to drive incoherent nuclear spin transitions via the Korringa process[27]; thus our previous analysis indicates that incoherent transitions driven by the electron current are very unlikely, and the nuclear spin dynamics are expected to be dominated by the Weger processes[28] as in the case of $TbPc_2$.[¡Error! Marcador no definido.] As seen in Figure 3 lower panel, this is not the case and thus the electronuclear spin states should be able to survive in the (infinitesimal voltage) conduction regime.

6: Extended discussion on hyperfine structure and anticrossings

There is a striking difference between the anticrossings in the case of $[V(\alpha\text{-}C_3S_5)_3]^{2-}$ compared with those of $TbPc_2$ (see fig X, fig SX). In the case of $TbPc_2$, for every electronuclear doublet there is an anticrossing, and the detection of these states happens due to a conductance jump at the anticrossing magnetic field. In contrast, in $[V(\alpha\text{-}C_3S_5)_3]^{2-}$ one of the crossings is allowed. This is a fundamental consequence of the different natures of the anticrossings in the two systems, which in $[V(\alpha\text{-}C_3S_5)_3]^{2-}$, a Kramers electronic spin, are purely from hyperfine origin and in $TbPc_2$ result from extradiagonal terms in the electronic spin Hamiltonian that cause tunneling splitting even before considering the nuclear spin.

The lack of an anticrossing will have consequences in the operating procedure, since it will not be possible to directly detect a transition between the electronuclear spin states (I,S)=(+3/2,+1/2),(-3/2,-1/2). These can be thought of as a "dark" doublet, analogous to the so-called "dark" states that cannot be directly detected in optical experiments. As we will see, in this case the lack of anticrossing does not impede indirect detection, thanks to the extraordinarily large tunneling splitting, which for vanadium complexes is at least three orders of magnitude larger compared with the case of $TbPc_2$. We need to recall that the detection happens via a Landau-Zener transition, described by the following equation.

$$P_{QTM} = 1 - \exp\left[ \frac{-\pi \cdot \Delta^{2}}{4\hbar g_s \mu_B \left| S_z \right| \mu_0 \frac{dH_z}{dt}} \right]$$

, where $\frac{dH_z}{dt}$ is the magnetic field sweep , $S_z = \pm\frac{1}{2}$ is the projection of the electronic spin, $\Delta$ is the tunnel splitting and $P_{QTM}$ is the tunnel probability.

Thus, an increase in three orders of magnitude of $\Delta$ compared with the experiments done on $TbPc_2$ means the corresponding transition probability is expected to be 1 for all practical field sweep rates. In turn, this means that not detecting a jump in conductance will be equivalent to measuring the nuclear spin state in the "dark" doublet.

The influence of the hyperfine coupling and the quadrupolar interaction also deserve attention, since they determine both the position and the probability of the anticrossing transitions (although, as just explained, the latter will be P=1 for all practical purposes). Two examples of hyperfine energy level structures can be seen in Figure S7, for the following parameter sets: (1: [$A_{\parallel}$=-340MHz, $A_{\perp}$=46MHz, $e^2qQ$=5.6MHz , $g_{\parallel}$=1.958, $g_{\perp}$=1.980] simplified from [$A_x$=-348Mhz, $A_y$=-310Mhz, $A_z$=46MHz, $e^2qQ$=5.6MHz, $g_x$=1.959, $g_y$=1.958, $g_z$=1.980] 2: [$A_{\parallel}$=418MHz, $A_{\perp}$=132MHz, $e^2qQ$=5.6MHz , $g_{\parallel}$=1.972, $g_{\perp}$=1.989 , ref 21, 22) corresponding to ($[V(\alpha\text{-}C_3S_5)_3]^{2-}$, $[VO(\alpha\text{-}C_3S_5)_2]^{2-}$. Note that the relation $A_{\parallel} > A_{\perp}$ is critical to obtain a confortable separation between the anticrossings in terms of magnetic field, which makes the experiment much less demanding.

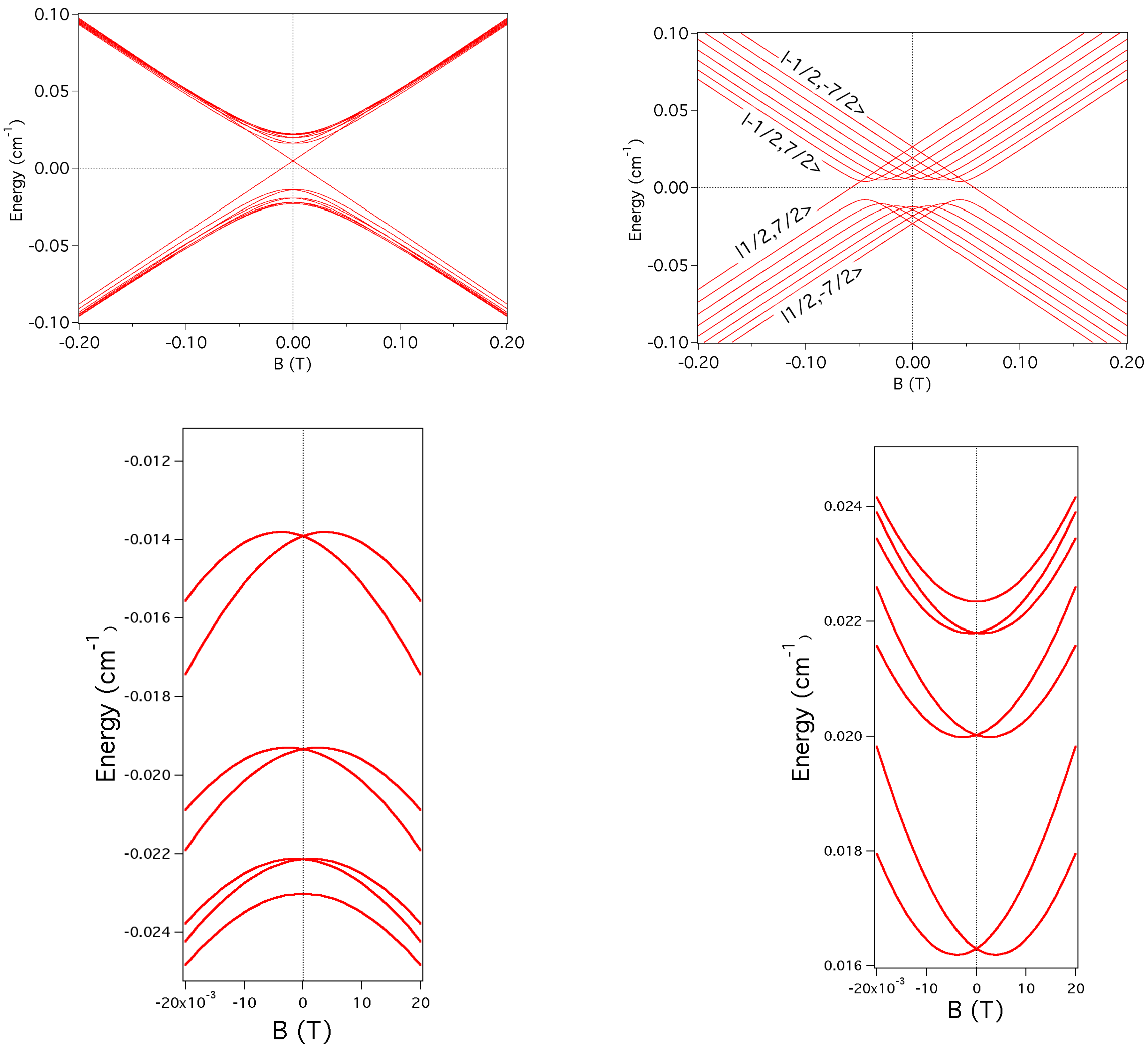

**Figure S7**: Evolution of the hyperfine structure (electronuclear spin energy levels) with an applied field for a vanadium complex **1** (left upper panel) and an analogous vanadyl complex **2** (right upper panel). While the same numbers of anticrossings are found, the tunneling splitting is larger for vanadium, but the field at which the anticrossings occur is larger for vanadyl. For clarity, zooms of the anticrossings for the vanadium complex are included in the lower panels, which all happen at a window of magnetic fields between -5mT and +5mT. Since the labeling of electronuclear states at varying fields is more complicated in this case, output files for SIMPRE are included as supplementary files.

Of course there is an important practical limitation if all transitions occur within a very small range of magnetic field: the more separated transition are, the easier it is to distinguish between different electronuclear doublets. However, here we need to consider that there will necessarily be an important difference between the hyperfine parameters that can be determined experimentally and the ones acting on a molecule contacted by two electrodes. In Figure 3 (down) of the main text, for example, one can see the symmetry breaking between the three ligands in the distribution of the electronic spin density. This limits our ability to predict experimentally relevant parameters such as the position and width of the anticrossings. Note for example that in the experiments that have been carried out on $TbPc_2$ the repetition of the experiment with different samples results in variations in a factor of 0.425 in the anticrossing width (Sample A: 0.34μK, Sample B: 0.8μK).

The ideal molecule would have the same conducting properties we have found (strongly spin-dependent conductance at zero gate voltage, no interference between the conducting channel and the electronic or nuclear spins, large tunneling splitting) but anticrossings that are more separated in terms of magnetic field. Analysis of EPR data in the literature point towards vanadyl complexes for the latter purpose, although there is by no means a unique solution. The well-characterized clock transitions in $HoW_{10}$ also display a large tunneling splitting and are sufficiently separated in magnetic field, so it this POM could be effectively contacted and presented a sufficient spin-dependent conductance it would also constitute an ideal solution.

## 7: References

[1] A. R. Rocha, V. Garcia-Suarez, S. W. Bailey, C. J. Lambert, J. Ferrer, S. Sanvito, *Nat. Mater.* 2005, **4**, 335. Ibid., *Phys. Rev. B* 2006, **73**, 085414. I. Rungger, S. Sanvito, *Phys. Rev. B* 2008, **78**, 035407.

[2] M. Soler, E. Artacho, J. D. Gale, A. Garcia, J. Junquera, P. Ordejon, and D. Sanchez Portal, *J. Phys. Condens. Matter*, 2002, **14**, 2745.

[3] H. Sellers, A. Ulman, Y. Shnidman, J. E. Eilers, *J. Am. Soc. Rev.* 1993, **115**, 9389.

[4] D.M. Ceperley, B. J. Alder, *Phys. Rev. Lett.* 1980, 45, 566.

[5] (1) C. D. Pemmaraju, T. Archer, D. Sánchez-Portal, S. Sanvito. *Phys. Rev. B: Condens. Matter.* 2007, 75, 045101-045116. (2) A. Filippetti, C. D. Pemmaraju, S. Sanvito, P. Delugas, D. Puggioni, V. Fiorentini, *Phys. Rev. B,* 2011, **84**, 195127-195149 (3) C. Toher, A. Filippetti, S. Sanvito, K. Burke, *Phys. Rev. Lett.*, 2005, **95**, 146402-146406.

[6] (1) R. B. Pontes, A. R. Rocha, S. Sanvito, A. Fazzio, J. R. Da Silva, *ACS Nano.*, 2011, **5**, 795-804. (2) W. R. French, C. R. Iacovella, I. Rungger, A. M. Souza, S. Sanvito, P. T. Cummings, *J. Phys. Chem. Lett.* 2013, **4**, 887-891. (3) W. R. French, C. R. Iacovella, I. Rungger, A. M. Souza, S. Sanvito, P. T. Cummings, *Nanoscale.* 2013, **5**, 3654-3663.

[7] C. Toher, S. Sanvito, *Phys. Rev. B,* 2008, **77**, 155402.

[8] N. Troullier, J. L. Martins, *Phys. Rev. B* 1991, **43**, 1993.

[9] M. Büttiker, Y. Imry, R. Landauer, S. Pinhas, *Phys. Rev. B,* 1985, **31**, 6207.

[10] S. Sanvito, *Ab-initio methods for spin-transport at the nanoscale level.* In Handbook of computational nanotechnology. **Vol 5**. American scientific pubishers, California 2004. Also available at arXiv:cond-mat/0503445 and references herein.

[11] Kepenekian, M.; Gauyacq, J.P.; Lorente, N. *J. Phys. Condens. Matter*, 2014, **26**, 104203

[12] McAdams, S. G.; Ariciu, A. M.; Kostopoulos, A. K.; Walsh, J. P. S.; Tuna, F. *Coordination Chemistry Reviews* **2017**, doi:10.1016/j.ccr.2017.03.015.

[13] AlDamen, M.; Clemente Juan, J. M.; Coronado, E.; Martí-Gastaldo, C.; Gaita-Ariño, A. *Journal of the American Chemical Society* **2008**, *130* (28), 8874. (2) (1) AlDamen, M.; Cardona-Serra, S.; Clemente Juan, J. M.; Coronado, E.; Gaita-Ariño, A.; Martí-Gastaldo, C.; Luis, F.; Montero, O. *Inorg. Chem* **2009**, *48* (8), 3467.

[14] Shiddiq, M.; Komijani, D.; Duan, Y.; Gaita-Ariño, A.; Coronado, E.; Hill, S. *Nature* **2016**, *531* (7594), 348.

[15] Cardona-Serra, S.; Clemente Juan, J. M.; Coronado, E.; Gaita-Ariño, A.; Suaud, N.; Svoboda, O.; Bastardis, R.; Guihéry, N.; Palacios, J. J. *Chem-Eur J* **2014**, *21* (2), 763.

[16] Lehmann, J.; Gaita-Ariño, A.; Coronado, E.; Loss, D. *Nature Nanotechnology* **2007**, *2* (5), 312.

[17] Jiang, S.-D.; Liu, S.-S.; Zhou, L.-N.; Wang, B.-W.; Wang, Z.-M.; Gao, S. *Inorg. Chem* **2012**, *51* (5), 3079.

[18] Martinez-Perez, M. J.; Cardona-Serra, S.; Schlegel, C.; Moro, F.; Alonso, P. J.; Prima-Garcia, H.; Clemente Juan, J. M.; Evangelisti, M.; Gaita-Ariño, A.; Sesé, J.; Van Slageren, J.; Coronado, E.; Luis, F. *Physical Review Letters* **2012**, *108*, 247213.

[19] Cardona-Serra, S.; Clemente Juan, J. M.; Coronado, E.; Gaita-Ariño, A.; Camón, A.; Evangelisti, M.; Luis, F.; Martinez-Perez, M. J.; Sesé, J. *Journal of the American Chemical Society* **2012**, *134*, 14982.

[20] Sherif, S.; Rubio-Bollinger, G.; Pinilla-Cienfuegos, E.; Coronado, E.; Cuevas, J. C.; Agrait, N. *Nanotechnology* **2015**, *26* (29), 291001.

[21] Zadrozny, J. M.; Niklas, J.; Poluektov, O. G.; Freedman, D. E. *ACS Central Science* **2015**, *9*, 488.

[22] Yu, C.-J.; Graham, M. J.; Zadrozny, J. M.; Niklas, J.; Krzyaniak, M. D.; Wasielewski, M. R.; Poluektov, O. G.; Freedman, D. E. *Journal of the American Chemical Society* **2016**, *138* (44), 14678.

[23] Bader, K.; Winkler, M.; Van Slageren, J. *Chem. Commun.* **2016**, *52*, 3623.

[24] Atzori, M.; Tesi, L.; Morra, E.; Chiesa, M.; Sorace, L.; Sessoli, R. *Journal of the American Chemical Society* **2016**, *138* (7), 2154.

[25] Sproules, S. Progress in Inorganic Chemistry, **2014**; *2*, 1–144.

[26] Cardona-Serra, S. et al. Work in preparation.

[27] Korringa, J. *Physica* **1950**, *16*, 601– 610.

[28] Weger, M. *Phys. Rev.* **1962**, *128*, 1505–1511.